\documentclass[a4paper,11pt]{article}

\usepackage{amsmath}
\usepackage{amssymb}
\usepackage{amsthm}
\usepackage{indentfirst}
\usepackage{graphicx}
\usepackage[utf8]{inputenc}

\usepackage{jheppub}
\hypersetup{colorlinks=true, linkcolor=blue, urlcolor=blue, citecolor=blue, linktocpage=true}

\newcommand{\g}{\mathfrak{g}}

\renewcommand{\L}{\mathcal{L}}
\newcommand{\LCDM}{$\Lambda$CDM }

\begin{document}

\title{Emergent Planck mass and dark energy from affine gravity}

\author[a,b]{I. Kharuk}

\affiliation[a]{Moscow Institute of Physics and Technology,\\
Institutsky lane 9, Dolgoprudny, Moscow region, 141700, Russia}
\affiliation[b]{Institute for Nuclear Research of the Russian Academy of Sciences,
\\ 60th October Anniversary Prospect, 7a, Moscow, 117312, Russia}

\emailAdd{ivan.kharuk@phystech.edu}

\preprint{INR-TH-2020-005}

\abstract{We introduce a novel model of affine gravity, which implements the no--scale scenario. Namely, in our model the Planck mass and Hubble constant emerge dynamically, through the mechanism of spontaneous breaking of scale--invariance. This naturally gives rise to the inflation, thus introducing a new inflationary mechanism. Moreover, the time direction and non--degenerate metric emerge dynamically as well, which allows one to consider the usual General Relativity as an effective theory. We show that our model is phenomenologically viable, both from the perspective of the direct tests of gravity and cosmological evolution.}

\maketitle

\section{Introduction}

The power of General Relativity (GR) comes from the fact that it combines profound physical ideas with elegant mathematical apparatus. By generalizing the latter, one may hope to gain a deeper insight into gravity. One of the possible routes in this direction is known as the affine gravity \cite{einstein1923theory,carmichael1925eddington,schrodinger1943general, schrodinger1943union, schrodinger1956space,schrodinger1944affine, kijowski1978new,ferraris1981general, liebscher1988purely}. The idea of this approach is based on the fact that any gravitational theory must include the connection field, but not (necessarily) the metric. Namely, in affine gravity the only independent field is the connection, while the metric is introduced as the momentum, canonically conjugated to the connection. This suggests a new look on the independent variables in gravity, which, for example, might be important for its quantization. However, it is known that affine gravity is phenomenologically equivalent to GR with a non--zero cosmological constant term \cite{fradkin1985quantum,ortin2004gravity,Magnano:1995pv}. In this context, the standard affine gravity does not provide new insights into the physics of gravity.

The aim of this paper is to introduce a novel model of affine gravity, which we dub scalar--affine gravity. In comparison with the standard affine gravity and GR, this model has the following features.

First, scalar--affine gravity is scale--invariant and realizes the no--scale scenario of \cite{minkowski1977spontaneous,zee1979broken, fujii1982origin,shaposhnikov2009scale, Blas:2011ac,kubo2019planck}. Namely, we start with a Lagrangian including only dimensionless constants. Then all physically important quantities, including Planck mass and Hubble parameter, emerge dynamically through the mechanism of spontaneous symmetry breaking (SSB). In particular, this naturally gives rise to the inflation. None of the previously considered models, including Born--Infeld type gravities, have these features.

Second, in scalar--affine gravity the vacuum solution is a de--Sitter--like space. Specifically, the metric is the same as in the de--Sitter space, but the background connection gets modified due to non--metricity. Such configuration of the vacuum is achieved by introducing a scalar field non--minimally coupled to gravity. Remarkably, the scalar field effectively acts as dark energy, thus providing an alternative to the cosmological constant term.

Finally, in our model the time direction and the metric emerge dynamically as well. Namely, we start with a general notion of a 4--dimensional manifold with the affine connection. Then, due to SSB, there appears a non--degenerate metric and one of the directions becomes distinguishable, playing the role of the time direction. In this context, our model realizes the idea of \cite{Borisov:1974bn,cutler1987new} that the metric can appear as a field breaking the general covariance down to the Poincare subgroup. We strengthen this approach by showing that the corresponding SSB also induces spontaneous origination of the time direction.

The model we are going to introduce has several drawbacks, which we discuss in the main part of the paper. However, we believe that the mentioned features of the scalar--affine gravity can renew the interest in the study of such models.

The paper is organized as follows. In section \ref{sec:2} we make an overview of affine gravity and introduce notations. In section \ref{sec:3} we present our model and discuss its features. In section \ref{sec:4} we show that our model is phenomenologically viable, both from the perspective of the direct tests of gravity and cosmological evolution. Finally, in section \ref{sec:5} we discuss the results and possible ways of generalizing the model.

\section{Affine gravity}
\label{sec:2}

\subsection{Overview of affine gravity}
\label{sec:2-1}

Depending on independent variables, all gravitational theories can be classified into three categories \cite{Magnano:1995pv}. 

The most often encountered are metric theories of gravity. In such theories the metric is the only source of gravity and the connection is fixed to be given by the Christoffel symbol of the metric. For example, GR belongs to this type of theories. 

Metric--affine theories, also known as the first--order (Palatini) formulations of gravity, form the second category. In this approach the metric and the connection are considered as independent variables. It is known that for a given $ f(R) $ theory, the metric and the metric--affine approaches yield different physical observables \cite{Sotiriou:2008rp}. 

The third class of gravitational theories constitute the affine theories. In such models the only source of gravity is the connection, while the metric is introduced as the momentum, canonically conjugated to the connection \cite{schrodinger1943general,kijowski1978new}. Affine theories of gravity were first formulated by Einstein and Eddington \cite{einstein1923theory, carmichael1925eddington}, further developed by Schrodinger \cite{schrodinger1943general,schrodinger1943union, schrodinger1956space,schrodinger1944affine}, and later by Ferraris and Kijowski \cite{kijowski1978new,ferraris1981general}. 

In affine gravity, the only tensor at hand is the Riemann curvature tensor,
\begin{equation}
R^\sigma_{\mu\nu\rho} = 2\partial_{[\mu}\Gamma^\sigma_{\nu]\rho} + 2\Gamma^\sigma_{[\mu|\lambda|} \Gamma^\lambda_{\nu]\rho} \;.
\end{equation}
Here and further we use parentheses and square brackets to denote, accordingly, symmetrization and antisymmetrization in the corresponding indices with proper weights. We also fix the dimension of the manifold to 4 and prescribe Greek letters to take values from 0 to 3. The contractions of the Riemann tensor give rise to the Ricci tensor,
\begin{equation}
R_{\mu\nu} \equiv R^\sigma_{\mu\sigma\nu} = 2\partial_{[\mu}\Gamma^\sigma_{\sigma]\nu} + 2\Gamma^\rho _{[\mu|\lambda|}\Gamma^\lambda_{\rho]\nu} \;,
\end{equation}
and to the homothetic curvature  tensor $ S_{\mu\nu} $,
\begin{equation} \label{homothetic_curv}
S_{\mu\nu} \equiv R^\sigma_{\mu\nu\sigma} = 2\partial_{[\mu}\Gamma_{\nu]\sigma}^\sigma \;.
\end{equation}
In the case of symmetric connection, the homothetic curvature coincides with the antisymmetric part of Ricci tensor, and it is an independent field otherwise. These are the building blocks of affine theories of gravity.

Throughout the paper we assume that the connection is torsionless and, for the reasons to become clear shortly, make use only of the symmetric part of the Ricci tensor. Then, in the absence of additional fields, the only diffeomorphism--invariant Lagrangian, up to multiplication by an arbitrary constant, reads
\begin{equation} \label{affine_classic}
\L =\sqrt{-\det \left( R_{(\mu\nu)}(\Gamma) \right)} \;.
\end{equation}
This is the Lagrangian of the simplest affine gravity model.

As Lagrangian above has a square root form, affine gravity can be considered as a subclass of Born--Infeld gravity \cite{jimenez2018born}. However, in most theories of Born--Infeld type the metric is considered as an independent primary field. As the question of identifying proper independent degrees of freedom might be of crucial importance (for example, for constructing quantum gravity), we prefer to distinguish affine gravity as a separate class.

There are two main ways of showing that affine gravity is equivalent to GR in the first--order formalism. The first approach employs the fact that in the first--order formalism GR's Lagrangian does not include derivatives of the metric. This allows one to consider the metric as a Lagrange multiplier and integrate it out. In the case of non--zero cosmological constant the resulting Lagrangian is that of the affine gravity. Indeed, consider GR with a non--zero cosmological constant term,
\begin{equation}
\L_{\Lambda} = \sqrt{-g} \left( M^2_{pl} R + 2\Lambda \right) \;,
\end{equation}
where $ M_{pl} $ is the Planck mass. Then by taking the determinant of Einstein's equations one gets
\begin{equation} \label{lagr_integr_metric}
\sqrt{-g}\left( M^2_{pl}R + 2\Lambda \right) = 2 \frac{M_{pl}^4}{\Lambda}\sqrt{-\text{det}R_{( \mu\nu ) } } \;,
\end{equation}
which proves the equivalence.

The second way exploits the definition of the metric as the momentum,  canonically conjugated to the connection. We will demonstrate this approach on a concrete example, closely following \cite{kijowski1978new}. Consider the Lagrangian
\begin{equation} \label{lagr_kij}
\L_{ex} = \frac{2}{m^2 \psi^2} \sqrt{-\det \left( \partial_\mu \psi \partial_\nu \psi - M_{pl}^2 R_{(\mu\nu)} \right)} \;,
\end{equation}
where $ m^2 $ is some constant and $ \psi $ is a scalar field. Let us introduce $ \g^{\mu\nu\rho}_\lambda $ as the momentum, canonically conjugated to the connection,
\begin{equation}
\label{def_momentum}
\g^{\mu\nu\rho}_\lambda \equiv \frac{\delta \L}{\delta \partial_\rho \Gamma^\lambda_{\mu\nu}} \;.
\end{equation}
Further, by introducing $ \g^{\mu\nu} $ as
\begin{equation}
\label{def_metric}
\g^{\mu\nu} \equiv \frac{\delta \L}{\delta R_{(\mu\nu)}}  \;,
\end{equation}
one can express $ \g^{\mu\nu\rho}_\lambda $ via $ \g^{\mu\nu} $ only,
\begin{equation}
\g^{\mu\nu\rho}_\lambda =  \delta^{(\mu}_\lambda \g^{\nu)\rho} - \delta^\rho_\lambda \g^{\mu\nu}  \;.
\end{equation}
Notice that $ \g^{\mu\nu} $ is automatically symmetric in its indices due to the symmetrization of $ R_{\mu\nu} $ in the Lagrangian and that its mass dimension equals two.  

Lagrangian (\ref{lagr_kij}) is a function of the connection, field $ \psi $, and their derivatives. Then, since the derivatives of the connection enter Lagrangian (\ref{affine_classic}) only via the Ricci tensor, its full differential can be written as
\begin{equation} \label{dif_lagr}
d \L = J^{\mu\nu}_\lambda d\Gamma^\lambda_{\mu\nu} + \g^{\mu\nu} d R_{(\mu\nu)} + J d \psi + P^\mu d \partial_\mu \psi
\end{equation}
for some $ J^{\mu\nu}_\lambda, ~ J $ and $ P^\mu $. To demonstrate the equivalence with GR, we perform the Legendre transformation by contact--deforming Lagrangian (\ref{lagr_kij}) in the gravitational sector,
\begin{equation} \label{contact_deformed}
L = \L - \g^{\mu\nu}R_{(\mu\nu)} \;.
\end{equation}
In this ``Hamiltonian'' formulation of gravity, $ \g^{\mu\nu} $ should be considered as an independent field, while $ R_{(\mu\nu)} $ as a function of $ \g^{\mu\nu} $ and $ \Gamma^\lambda_{\mu\nu} $. The full differential of (\ref{contact_deformed}) is thus
\begin{equation} \label{diff_hamilt}
dL = J^{\mu\nu}_\lambda d\Gamma^\lambda_{\mu\nu} -  R_{(\mu\nu)} d \g^{\mu\nu} + J d \psi + P^\mu d \partial_\mu \psi \;.
\end{equation}
By definition, this implies
\begin{equation} \label{def_hamilt_ricci}
R_{(\mu\nu)} = - \frac{\delta L}{\delta \g^{\mu\nu}} \;.
\end{equation}
Then, after introducing the usual metric $ g_{\mu\nu} $ as
\begin{equation} \label{metr_redef_calssic}
\g^{\mu\nu} \equiv - M^2_{pl} \sqrt{-g}g^{\mu\nu} \;
\end{equation}
and employing the formula for the compound derivative, eq. (\ref{def_hamilt_ricci}) can be cast to the form of the Einstein equations \cite{kijowski1978new}. 

It remains to notice that the variation of the action with respect to $ \Gamma^\lambda_{\mu\nu} $ yields the metric--compatibility condition,
\begin{equation} \label{metr_comp}
\nabla_\lambda \left( \sqrt{-g} g^{\mu\nu} \right) = 0 \;,
\end{equation}
and that $ \psi $ obeys the Klein--Gordon equation for a scalar field with mass $ m^2 $. Thus, Lagrangian (\ref{lagr_kij}) can be replaced by the equivalent one --- of a usual massive field coupled to the standard GR.

\subsection{Perspectives and problems of affine gravity}
\label{sec:2-2}

If affine gravity is equivalent to GR, one may wonder what are the benefits of employing the affine formalism. We believe that there are two major benefits.

First, affine gravity illustrates that the metric might not be a fundamental field \cite{kijowski1978new}. Instead, in this approach it appears as the momentum, conjugated to the connection, whose equations of motion are precisely the Einstein equations. As we discuss in section \ref{sec:5-2}, this change of perspective on the fundamental variables allows one to approach the gravity as a gauge theory of the connection. 

The point above also makes affine gravity more natural than GR from the geometrical perspective. Indeed, the notion of the connection is of primary importance for comparing vectors at different points. On the other hand, the metric, although it has a clear physical meaning, is always an additional structure. In contrast to GR, affine gravity exploits only the primary geometrical object --- the connection --- and by eq. (\ref{def_metric}) unambiguously defines the physical metric.

The second reason is that the affine gravity allows for a wide range of modifications. For example, one can choose not to perform the symmetrization of $ R_{\mu\nu} $'s indices in Lagrangian (\ref{affine_classic}), thus obtaining a non--symmetric metric. Earlier this was considered as a possibility to incorporate the electromagnetism and strong interactions \cite{einstein1923theory,schrodinger1943union} into the unified geometrical model. However, the studies showed that this and similar ideas cannot be successfully implemented. Nonetheless, possible extensions of affine gravity are a subject of an ongoing work \cite{ferraris1982unified,Krasnov:2007ei,Krasnov:2011pp, Poplawski:2007rx,Poplawski:2012bw}, including newly introduced polynomial affine gravity \cite{Castillo-Felisola:2015cqa,Castillo-Felisola:2019wcs} and affine--based models of the inflation \cite{azri2018induced,azri2018cosmological}.

At this point we would also like to make contact with the idea that gravity should be described by a spin--2 field with spin--2 gauge invariance or general covariance \cite{weinberg1964photons,deser1970self, boulware1975classical,wald1986spin, cutler1987new}. An important remark is that in this approach the graviton does not have to be a fundamental particle. Instead, it might be some composite or even auxiliary field. Hence gravitational theories based on ideas of \cite{weinberg1964photons,deser1970self, boulware1975classical,wald1986spin, cutler1987new} should be considered as effective theories of gravity, and affine gravity is consistent with these ideas.

A major problem in affine gravity is that it does not allow for a unified description of the Standard Model and gravity. Namely, one can add the Lagrangian of the Standard Model to the Hamiltonian of the gravity, eq. (\ref{contact_deformed}). However, it is unknown how (and if) such full theory can be translated back to the Lagrange formulation of gravity, i.e., in the form similar to eq. (\ref{affine_classic}). In particular, it is unknown how one should introduce fermions in the Lagrange formulation.

\section{Scalar--affine gravity}
\label{sec:3}

Let us introduce the scalar--affine gravity. As the name of the model suggests, we introduce a scalar field $ \varphi $, a scalar tensor density of weight $ w $. Accordingly, its covariant derivative reads
\begin{equation} \label{def_covar_phi}
\nabla_\mu \varphi = \partial_\mu \varphi - w \Gamma^\sigma_{\mu\sigma} \varphi \;.
\end{equation} 
Note that it introduces non--minimal coupling between $ \varphi $ and gravity. The Lagrangian of our model is
\begin{equation} \label{model_lagr}
\L = \sqrt{ -\det L_{\mu\nu} } + \alpha \varphi^{w^{-1}} \,, ~~~~ L_{\mu\nu} = R_{(\mu\nu)} + \frac{c}{2} \frac{\nabla_\mu \varphi \nabla_\nu \varphi}{\varphi^2} \;,
\end{equation}
where $ \alpha $ and $ c $ are some constants. By redefining $ \varphi $ one can set $ w = 1 $ and $ \alpha = 1 $, which is used from here onwards.\footnote{For a general $ w $ this procedure may require redefining $ \varphi $ to be a purely imaginary field. In this case under $ \varphi $'s value one should understand its absolute value.}

Let us discuss this Lagrangian. First, we note that in order to compensate the transformation of the volume element under an arbitrary change of coordinates, the Lagrangian must be a scalar density of weight +1. This implies that $ L_{\mu\nu} $ must be an absolute tensor, i.e., a density of zero weight. Then the only allowed first--order $ \varphi $'s kinetic term is the one introduced in Lagrangian (\ref{model_lagr}). 

Further we note that our model is scale--invariant --- it does not include dimensionfull constants. In terms of symmetry transformations, it is invariant under the transformation
\begin{equation} \label{SI_affine}
\Gamma^\rho_{\mu\nu}(x) \rightarrow \lambda \Gamma^\rho_{\mu\nu}(\lambda x) \,, ~~~~ \varphi(x) \rightarrow \lambda^4 \varphi(\lambda x) \;,
\end{equation} 
where $ \lambda $ is some constant. These transformation rules are direct analogues of the standard, metric--based formulation of the scale--invariance. Indeed, keeping the definition of the metric as the momentum, canonically conjugated to the connection, eq. (\ref{def_metric}), under replacement (\ref{SI_affine}) metric $ \g^{\mu\nu} $ transforms as
\begin{equation}
\g^{\mu\nu}(x) \rightarrow \lambda \g^{\mu\nu}(\lambda x) \;.
\end{equation}
Together with $ \varphi $'s transformation they form the standard definition of scale--invariance.  

Lagrangian (\ref{model_lagr}) is not the most general one that can be written down using the tensors at hand. Possible additional terms fall into two categories.   

First, one can add a second--order derivative of $ \varphi $ to $ L_{\mu\nu} $,
\begin{equation} \label{add_1}
L_{\mu\nu} \rightarrow L_{\mu\nu} + c_1 \nabla_\mu \frac{\nabla_\nu \varphi}{\varphi}
\end{equation} 
for some constant $ c_1 $. This term, however, violates the equivalence with GR. Indeed, since it contains derivatives of the connection, the contact deformed Lagrangian, eq. (\ref{contact_deformed}), cannot be written in the form, similar to (\ref{diff_hamilt}). This breaks down the identities (\ref{def_hamilt_ricci}), which are, in fact, Einstein equations. As there is no phenomenological evidence for considering such modified theories of gravity, we forbid terms like (\ref{add_1}) in the Lagrangian.

Second, one can introduce more root structures into the Lagrangian, like
\begin{equation} \label{add_root_1}
\L_{add} = \sqrt{- \det \frac{c_2}{2} \frac{\nabla_\mu \varphi \nabla_\nu \varphi}{\varphi^2} } \;,
\end{equation}
where $ c_2 $ is some constant. We expect that in a complete theory of affine gravity some symmetry will restrict the Lagrangian to be a certain function of $ R_{\mu\nu} $ and other fields. In particular, it will forbid the additional square root terms. At present we do not know the corresponding symmetry but assume that the terms like one above are forbidden. Finally, one can also consider the presented model as an illustrative one, showing the possibilities of the approach. 

Theories, most similar to the introduced one, were considered in \cite{jimenez2018born,vollick2005born}. However, none of them are scale--invariant or include coupling between the matter fields and the affine connection. In this perspective, scalar--affine gravity represents a new model, not studied anywhere previously.

Let us now qualitatively discuss the dynamics of our model. The $ \varphi^2 $ term appearing in the denominator of $ \varphi $'s kinetic term initiates spontaneous breakdown of scale invariance. Indeed, $ \varphi $'s equation of motion (EqM) read 
\begin{equation} \label{phi_eqM_gen}
-\nabla_\lambda \left( \g^{\mu\nu} \frac{\delta L_{\mu\nu}}{\delta \nabla_\lambda \varphi} \right) + \g^{\mu\nu} \frac{\delta L_{\mu\nu}}{\delta\varphi} + 1 = 0 \;.
\end{equation}
Because this equation contains terms with negative power of $ \varphi $, as well as a constant term, on the solution $ \varphi \neq 0\,, ~ \varphi \neq \infty $. In turn, this forces $ \g^{\mu\nu} \neq 0 $ as well. This is the mechanism for the emergence of non--degenerate metric and dimensionfull parameters in our model.

To study the dynamics of the model, we replace Lagrangian (\ref{model_lagr}) by an equivalent one,
\begin{equation} \label{model_lagr_metric}
\L_{equiv} = -4\sqrt{-\text{det}\tilde{\g}^{\mu\nu}} +\tilde{\g}^{\mu\nu} L_{\mu\nu} + \varphi \;.
\end{equation}
Here it is understood that $ \tilde{\g}^{\mu\nu} $ and the connection are independent variables. The two Lagrangians are equivalent as by integrating $ \tilde{\g}^{\mu\nu} $ out from (\ref{model_lagr_metric}) one arrives at the initial Lagrangian. In particular, since $ \tilde{\g}^{\mu\nu} $'s EqM coincide with that of $ \g^{\mu\nu} $, we will omit tildes over $ \tilde{\g}^{\mu\nu} $ further. Thus, by introducing $ \g^{\mu\nu} $ as an independent field we get rid of the square root structure and arrive at a commonly known first--order formulation of gravity.

As we know, on the solution $ \varphi \neq 0 $. This allows us to redefine the fields as follows,
\begin{equation} \label{field_reparam_init}
\varphi = \varphi_0^4 e^{\pi} \,,  ~~~~~ \g^{\mu\nu} = - \varphi_0^{2}\sqrt{-g}g^{\mu\nu} \;.
\end{equation}
where $ \varphi_0 $ is some constant of unit mass dimension and $ g^{\mu\nu} $ is a non--degenerate symmetric absolute rank 2 tensor. Now $ \pi $ can be thought of as the dilaton and $ g_{\mu\nu} $ as the metric. We will use this parametrization of the fields further.

It is convenient to start the analysis of the model by considering EqM in the gravitational sector. The variation of the action with respect to $ \Gamma^\lambda_{\mu\nu} $ yields
\begin{equation} \label{EqM_Connection}
\nabla_\lambda \g^{\mu\nu} - \delta^{ ( \mu}_\lambda\nabla_\sigma\g^{|\sigma|\nu ) } = c \g^{\sigma(\mu}\delta^{\nu)}_\lambda \frac{\nabla_\sigma\varphi}{\varphi} \;.
\end{equation}
Importantly, from this equation one gets
\begin{equation} \label{to_gauge}
\left( 1+\frac{c}{3} \right) \Gamma^{\sigma}_{\sigma\mu} = \partial_{\mu} \ln \sqrt{-g} + \frac{c}{3}\partial_\mu \ln \varphi \;.
\end{equation}
For a special value of $ c = -3 $ this reduces to
\begin{equation} \label{constr_phi}
\varphi = \varkappa \varphi_0 \sqrt{-g} \;,
\end{equation} 
where $ \varkappa $ is some constant. Thus, in this special case $ \varphi $ is not an independent degree of freedom, and the total number of degrees of freedom coincides with that in GR. This is our motivation for fixing $ c = -3 $. As we demonstrate below, this is consistent with $ \varphi $'s EqM and ensures that the Planck mass is a constant (as a function of time).

The solution of eq. (\ref{EqM_Connection}) reads
\begin{equation} \label{connect_via_metric}
\Gamma^\lambda_{\mu\nu} = \lbrace ^\lambda _{\mu\nu} \rbrace - \frac{1}{2} ( v_\mu \delta^\lambda_\nu + v_\nu \delta^\lambda_\mu - 3 g_{\mu\nu} g^{\lambda\rho}v_\rho ) \,, ~~~~ v_\mu \equiv \frac{\nabla_\mu\varphi}{\varphi} \;,
\end{equation}
where $ \lbrace ^\lambda _{\mu\nu} \rbrace  $ is the Christoffel symbol of the metric $ g_{\mu\nu} $. As we see, there is non--zero nonmetricity due to the coupling of $ \varphi $ to the connection.

Having established these facts, we move on to solving the Einstein equations. For this purpose we first obtain $ \varphi $'s energy--momentum tensor. From $ \varphi $'s EqM we know that $ \nabla_\mu \varphi \neq 0 $. Hence we can choose coordinates in which $ \nabla_\nu \varphi $'s direction coincides with the 0--axis,
\begin{equation} \label{anz_deriv}
\frac{\nabla_\mu\varphi}{\varphi} = \left( -2 v,0,0,0 \right) \;,
\end{equation}
where $ v $ is an arbitrary function of the coordinates. Let us denote by $ T_{\mu\nu}^\varphi ~ \varphi $'s energy--momentum tensor plus terms proportional to $ v^2 $ coming from the Einstein tensor due to non--metricity. Then one has
\begin{equation}
T_{\mu\nu}^\varphi = -3v^2g_{\mu\nu} \;. 
\end{equation}
We see that $ \varphi $ effectively acts as the dark energy. Further, by using the redefinition of the fields, eq. (\ref{field_reparam_init}), one sees that the determinant of $ \g^{\mu\nu} $ in Lagrangian (\ref{model_lagr_metric}) also acts as a cosmological constant term. Hence the solution is an Einstein manifold, with the metric of the form
\begin{equation} \label{anz_metr}
g_{\mu\nu} = \text{diag} \left( -1,a^2(t),a^2(t),a^2(t) \right) \;.
\end{equation}
Notice that the  $ -1 $ element of the metric correlates with $ \nabla_\mu \varphi $'s direction. This is a must since each of them defines a distinguishable direction, which must coincide. Thus, in our model the time direction emerges dynamically, with $ \nabla_\mu \varphi $ defining the proper time axis.

For the metric (\ref{anz_metr}) the non--zero components of the connection are
\begin{equation} \label{connection_vaccum}
\Gamma^0_{ij} = a^2 ( H+3v )\gamma_{ij} \,, ~~~ \Gamma^i_{0j} = ( H+v )\delta^i_j \,, ~~~ \Gamma^0_{00} = -v \,, ~~~ H \equiv \frac{a'}{a} \;,
\end{equation}
where Latin indices stay for spatial components, $ i =1,2,3 $, apostrophe denotes differentiation with respect to time,  and  $ \gamma_{ij} = a^{-2} g_{ij} $ is the flat metric. Then from the Einstein equations (or, equivalently, by varying the action with respect to $ \g^{\mu\nu} $) one gets 
\begin{subequations} 
\begin{align}  \label{eq_consist_Hprime}
H' &= 0 \,, \\
\label{eq_consist_param}
3(H^2+3Hv + &v') = 2 \varphi_0^2  \;,
\end{align}
\end{subequations}
Hence $ H $ is a constant while $ v $ might be time--dependent.

Consider now $ \varphi $'s EqM. After taking into account that $ \varphi $ follows the dynamic of the determinant of the metric, eq. (\ref{constr_phi}), it reads
\begin{equation} \label{solut_phi}
6 \left( 3Hv + v' \right) = - \varkappa \varphi_0^2 \;.
\end{equation}
Together with eq. (\ref{eq_consist_param}), they form a full set of equations of the scalar--affine gravity. Any of the dimensionfull parameters --- say, $ \varphi_0 $ --- can be chosen as a reference unit. Then these equations define the ratios of the other dimensionfull parameters to the reference one. Thus, scalar--affine gravity realizes the no--scale scenario. 

By comparing eq. (\ref{field_reparam_init}) and (\ref{metr_redef_calssic}) and applying the arguments of section \ref{sec:2}, one concludes that $ \varphi_0 $ is the Planck mass. Then we have\footnote{$ \varkappa = -4 $ corresponds to a special solution, for which $ H = 0 $ and $ v $ is time--dependent. We will not consider this case here.}
\begin{equation}
\left( \frac{H}{M_{pl}} \right)^2 = \frac{\varkappa + 4}{6} \;.
\end{equation}
As we demonstrate below, for the mass of elementary particles to be much smaller than the Planck mass, it must hold that $ \varkappa \ll 1 $. In this case the Hubble constant is of order of the Planck mass. This naturally gives rise to the inflation as a consequence of spontaneous breakdown of scale--invariance, which is accompanied by the emergence of the metric. This seems to be an interesting starting point for the ``birth'' of the Universe. However, at present, we do not know how to end the inflation and leave this question for future study.

Let us now discuss how one can add fields into the model on the example of a scalar field $ \chi $. Since we already have a scalar density field in the theory, $ \chi $ can be introduced as an absolute tensor. Then requiring the Lagrangian to be a sum of a square root term and a potential, the most general Lagrangian reads
\begin{equation}
\L = \sqrt{ -\det \left( L_{\mu\nu} + \frac{q}{2} \partial_\mu \chi \partial_\nu \chi \right) } +  \varphi (1 + \beta \chi^2) \;,
\end{equation}
where $ \beta $ is some constant and $ q = \pm 1 $. Assuming $ \beta>0 $ (thus $ \chi $ does not participate in the SSB), the equivalent first--order Lagrangian, with the metric as an auxiliary field, is
\begin{equation} \label{lagr_with_matter}
\L_{equiv} = -4\sqrt{-\text{det}\g^{\mu\nu}} +\g^{\mu\nu} L_{\mu\nu} + \varphi + \g^{\mu\nu} \frac{q}{2}  \partial_\mu \chi \partial_\nu \chi + \beta\varphi \chi^2 \;.
\end{equation}
Also notice that, unlike the usual fields in common field theories, $ \chi $'s mass dimension is zero. In order to normalize it conventionally, we redefine it as follows,
\begin{equation}
\chi \rightarrow \varphi_0^{-1} \chi \;.
\end{equation} 
Now $ \chi $'s mass dimension is unity and its kinetic term is properly normalized. Then from the Lagrangian one reads out $ \chi $'s mass,
\begin{equation}
m^2_{\chi} = \beta \varkappa M_{pl}^2.
\end{equation}
Assuming $ \beta $ is of order unity, elementary particles are much lighter than the Planck mass only in the regime $ \varkappa \ll 1$. 

\section{Phenomenological validity}
\label{sec:4}
 
\subsection{Linearized limit}
\label{sec:4-1}

Because of the non--zero non--metricity, scalar--affine gravity might not be phenomenologically viable even in the linearized limit. In appendix \ref{sec:A} we show that this is not the case --- EqM in all of the helicity sectors coincide with that of GR. Hence our model describes phenomenologically viable gravitational waves and reproduces Newton's law.

\subsection{Cosmology}
\label{sec:4-2}

As we have mentioned previously, our model does not provide means for ending the inflation. However, if we assume that the inflation has somehow finished, scalar--affine gravity gives rise to phenomenologically viable cosmology. Indeed, after SSB took place, one can introduce all fields of the Standard Model into the theory as the corresponding representations of the stability group of the metric. It remains to show that the standard conservation laws hold in our model. Below we prove this statement. 

Let us first discuss $ \Gamma^\lambda_{\mu\nu} $'s and $ \varphi $'s EqM in the presence of matter. We assume that matter fields do not couple to $ \varphi $'s covariant derivative or to the connection. This guarantees that the introduced earlier mechanism of spontaneous breakdown of scale--invariance is unaffected by the usual matter. Then $ \Gamma^\lambda_{\mu\nu} $'s EqM remain unchanged, with solution given by eq. (\ref{connect_via_metric}), and $ \varphi $ is given by eq. (\ref{constr_phi}) at any time. In particular, in the leading order in the magnitude of fields, $ \varphi $'s EqM is given by eq. (\ref{solut_phi}).

Further, as usually, we assume that the Universe can be described in the hydrodynamic limit and search for a dS--like solutions of the Friedmann equations. The latter read
\begin{subequations} \label{Fr_eq_gen}
\begin{align}
(00)&:~~~ 3v' + 9vH - 2\varphi_0^{\frac{c}{2}} + 3H^2 = 8 \pi M_{pl}^{-2} \rho \;, \\
(ii)&:~~~ 3v' + 9vH - 2\varphi_0^{\frac{c}{2}} + 3H^2 + 2H' = - 8 \pi M_{pl}^{-2} p \;,
\end{align}
\end{subequations}
where $ \rho \,,~  p $ are matter energy densities and pressure correspondingly. As it follows from eq. (\ref{solut_phi}), the first three terms on the left hand side of both of the equations are constants. Then by introducing ``dark energy'' density and pressure,
\begin{equation}
- \rho_{de} = p_{de} = 3v' + 9vH - 2\varphi_0^{\frac{c}{2}} \;,
\end{equation} 
Friedman equations get reduced to the usual ones in \LCDM cosmology. An immediate consequence of these equations is the usual conservation law in the matter sector,
\begin{equation} \label{cons_matter}
\rho' + 3H(p+\rho) = 0 \;.
\end{equation}
This implies that the cosmological evolution of matter densities as functions of $ a^2 $ are the same as in the \LCDM cosmology.

Interestingly, in our model the connection is time--dependent during the dust and radiation dominating epochs. For example, for the dust dominating epoch one gets
\begin{equation}
v = - \frac{\varkappa}{18} M_{pl}^2 t \;.
\end{equation}
Since the connection depends on $ v $, the latter also becomes time--dependent. Provided that $ \varkappa \ll 1 $, this has small influence on the cosmological evolution. Nonetheless, as fermions do couple to the connection, this might have observable phenomenological effects. We comment on this question in the next section.

We would like to end this section by noticing that scalar--affine gravity admits a Schwarzschild--like solution. Namely, the metric is the same as in the Schwarzschild solution, but, due to non--metricity, the background connection gets modified. This fact might have important phenomenological consequences. However, before studying them, we should first provide more solid ground for the foundation of our model. Correspondingly, we leave a detailed study of this question for future.
 
\section{Discussion}
\label{sec:5}

\subsection{Possible physical implications}
\label{sec:5-1}

Although scalar--affine gravity cannot account for all phenomenological data, it has some remarkable and promising features. First, as we have already mentioned, it realizes the no--scale scenario --- all physical quantities are generated dynamically through the SSB mechanism. Combined with the fact that scalar--affine gravity is phenomenologically viable, it might provide a foundation for a scale--invariant theory of gravity. In particular, the presence of the additional field $ \varphi $ might improve UV behavior of gravity.

Second, as we have shown, scalar--affine gravity naturally gives rise to the inflation. Namely, we do not need to postulate any specific form of the potential --- the scalar--density field $ \varphi $ effectively acts as dark energy, thus giving rise to the expanding Universe. Although our model does not provide means for ending the inflation, this problem, presumably, can be solved by extending the field content of the model.

Third, in our model the time direction and Lorentz--invariance (in the flat limit) emerge dynamically. In this perspective scalar--affine gravity is an alternative to the standard idea of the emergent Lorentz invariance \cite{nielsen1978beta,chadha1983lorentz, bednik2013emergent}, as well as to the Lorentz--violating gravity \cite{rubakov2004lorentz,dubovsky2004phases, rubakov2008infrared,Blas:2014ira}, including, in particular, the Horava gravity \cite{hovrava2009quantum,hovrava2009membranes}.

Finally, scalar--affine gravity features non--zero non--metricity --- although the metric is of the dS form, the connection is not canonical. In particular, since $ \Gamma^0_{00} \neq 0 $ and fermions couple to the connection, this implies that all fermions are massive, with the mass of the order $ v $. This provides us with an alternative to $ \nu $MSM mechanism for the generation of neutrino's masses \cite{asaka2005numsm}. However, to discuss this topic we first need to fully embed the Standard Model into our theory, including, in particular, the Higgs field.

In this context we would also like to note that the scalar--affine gravity does not provide means for explaining the signature of the metric. Namely, in Lagrangian (\ref{model_lagr}) one can change the minus sign before the determinant to the plus sign. In this case the metric will have Euclidean signature. It will be interesting to study whether there exists a mechanism for fixing the minus sign in Lagrangian (\ref{model_lagr}). This might be closely related to the idea of dynamical generation of Lorentzian signature
of the metric by introducing an additional scalar field \cite{Greensite:1992np, Elizalde:1992ug, Carlini:1993up}

\subsection{Gravity as a gauge theory of connection}
\label{sec:5-2}

One of the approaches to constructing a theory of gravity lies in formulating it as a usual gauge theory \cite{macdowell1977unified,chamseddine1978massive, chamseddine1977supergravity,stelle1980spontaneously, hehl1995metric}. Affine theories of gravity allow to approach this problem from a new perspective. Namely, one can try to construct gravitational theory as a gauge theory of the connection. Below we discuss this idea and how it might be used as a foundation of the scalar--affine gravity.

Since the basic field of the affine gravity is the connection, it is natural to remember Einstein's $ \lambda $--transformations \cite{einstein2003meaning},
\begin{equation} \label{lambda_connection}
\Gamma(x)^\lambda_{\mu\nu} ~ \rightarrow ~ \Gamma(x)^\lambda_{\mu\nu} + \partial_\mu \lambda(x) \delta^\lambda_\nu \;, 
\end{equation}
where $ \lambda(x) $ is an arbitrary function. Riemann curvature tensor is known to be invariant under such transformations. In fact, GR possesses even a larger symmetry, namely,
\begin{equation} \label{proj_inv}
\Gamma^\lambda_{\mu\nu} \rightarrow \Gamma^\lambda_{\mu\nu} + \xi_\mu \delta^\lambda_\nu \;, 
\end{equation}
where $ \xi_\mu $ is an arbitrary function \cite{Julia:1998ys,Dadhich:2010xa}. This symmetry is known as (a subgroup of) the protective invariance \cite{eisenhart2012non}. To reveal its geometrical meaning, consider some geodesic \cite{schrodinger1943union}. In affine geometry there is no notion of a predefined metric. Correspondingly, the equation of the geodesic can be defined only as a parallel transport of a vector, remaining parallel to itself,
\begin{equation} \label{geo_without_metric}
\frac{d^2 x^\mu}{dp^2} + \Gamma^\mu_{\sigma\rho}\frac{dx^\sigma}{dp}\frac{dx^\rho}{dp} = \chi(p)\frac{dx^\mu}{dp} \;,
\end{equation} 
where $ p $ is the parameter along the geodesic and $ \chi(p) $ is some function. Because of the form of the second term on the left hand side of the equation, only the symmetric part of the connection, which we denote as $ \Upsilon^\lambda_{\mu\nu} $, is relevant. Then a general transformation of the connection leaving a geodesic invariant (up to a reparametrization) reads \cite{schrodinger1943union,schrodinger1956space, eisenhart2012non}
\begin{equation}
\Upsilon^\lambda_{\mu\nu} ~ \rightarrow ~ \Upsilon^\lambda_{\mu\nu} + \delta^\lambda_\nu V_\mu + \delta^\lambda_\mu V_\nu \;,
\end{equation}
where $ V_\nu $ is an arbitrary function. These transformations are known as projective transformations and are wider than transformations (\ref{proj_inv}).

Projective symmetry might form the basis for formulating the scalar--affine gravity as a gauge--type theory. Indeed, in the absence of predefined notion of the metric, eq. (\ref{geo_without_metric}) is the only possible definition of a geodesic. If one considers geodesics as primary objects, all of their symmetries must be symmetries of the full theory as well. Correspondingly, we expect a general theory of scalar--affine gravity to incorporate the projective symmetry.  

To demonstrate why this idea is promising, consider Einstein's $ \lambda $--transformations for the trace of the connection,
\begin{equation}
\Gamma^\sigma_{\mu\sigma} ~ \rightarrow ~ \Gamma^\sigma_{\mu\sigma} + 4 \partial_\mu \lambda \;. 
\end{equation}
They mimic the transformation law of the $ U(1) $ 4--potential $ A_\mu $ under the $ U(1) $ gauge symmetry. The analogy goes further --- the homothetic curvature  tensor $ S_{\mu\nu} $, eq. (\ref{homothetic_curv}), has the same structure as the electromagnetic tensor $ F_{\mu\nu} \equiv \partial_\mu A_\nu - \partial_\nu A_\mu $. This suggests that the trace of the connection may be somehow identified with the 4--potential, and $ \lambda $--transformations with the corresponding gauge invariance \cite{borchsenius1976extension}. Such an extension might be of interest for extending the Standard Model, with $ U(1) $ as a symmetry of some hidden or yet unobserved sector \cite{ferraris1981general,ferraris1982unified, Poplawski:2007rx, Castillo-Felisola:2019wcs,Poplawski:2007ik, Krasnov:2017epi, borchsenius1976extension, Kharuk:2018ums}. 

In conclusion we would like to say that there is a number of open questions in scalar--affine gravity. Nonetheless, it provides new ways for approaching long--standing problems in gravity. We believe that this makes it worth considering and studying further. 

\acknowledgments

The author thanks V. Rubakov and A. Shkerin for useful discussions.

\appendix

\section{Linearized limit}
\label{sec:A}

To verify phenomenological validity of the scalar--affine gravity, we show that linearized limit of our model coincides with that of GR. We parametrize the fluctuations of the metric as follows,
\begin{equation}
g_{00} = \eta_{00} + h_{00} \,, ~~~~~
g_{ij} = a^2 ( \eta_{ij} + h_{ij} ) \,, ~~~~~ g_{0i} = h_{0i}  \;,
\end{equation}
where $ h_{\mu\nu} $ are fluctuations of the metric. In the rest of this section the sum is taken with respect to the flat metric $ \delta_{ij} $, and we do not distinguish upper and lower indices. We employ the $ 3+1 $ decomposition,
\begin{subequations} 
\begin{align}
h_{00} &= 2\Phi \,, \\
h_{0i} &= \partial_i Z + Z_i^T \,, \\
h_{ij} &= -2\Psi\delta_{ij} + 2\partial_i \partial_j E + \partial_{( i}W_{j)}^T + h_{ij}^{TT} \;,
\end{align}
\end{subequations}
where, as usual,
\begin{equation}
\partial_i Z_i^T = 0 \,, ~~~ \partial_i W_i^T = 0 \,, ~~~ \partial_i h_{ij}^{TT} = 0 \,, ~~~ h_{ii}^{TT} = 0 \;,
\end{equation}
and impose the gauge 
\begin{equation}
h_{0i} = 0 \;.
\end{equation} 

For our choice of constant $ c=-3 $, $ \varphi $ follows the dynamics of the determinant of the metric. Hence their fluctuations are the same. However, we also need to obtain the formula governing the fluctuations of $ \varphi $'s covariant derivative. Then for $ v_\mu $'s fluctuations, $ v_\mu = v_\mu^{vac} + u_\mu $, one gets the equation
\begin{equation}
-(a^3\partial_0 + 3a^2\partial_0 a) \left( u_0 + \frac{v}{2}(h+h_{00}) \right) + a \partial_i u_j = 0 \;,
\end{equation}
where $ h = h_i^i $. The solution of this equation reads
\begin{equation} \label{eq_scal_fluct}
u_0 = -\frac{v}{2}(h-h_{00}) - u \,, ~~~~ \dot{u} + 3Hu + a^{-2} \partial_i u_i = 0  \;.
\end{equation} 

Now we consider EqM for the perturbations of the metric. As it follows from eq. (\ref{connect_via_metric}), the quadratic terms in $ h_{\mu\nu} $ are suppressed provided that
\begin{equation} \label{cond_for_expand}
h_{\cdot \cdot} \ll 1 \,, ~~~~ v h_{\cdot \cdot} h_{\cdot \cdot} \ll \partial_{\cdot} h_{\cdot \cdot} \;,
\end{equation}
where dots stay for some indices. The first condition is standard. The second constraint appears due to non--zero non--metricity and its validity should be verified after obtaining the solution of EqM. 

We use Cadabra software \cite{peeters2018cadabra2,peeters2007introducing} for obtaining EqM for the perturbed metric. The result is that the equations of motions in spin--2 and spin--1 sectors are the same as in GR on the de Sitter background. Hence in our model gravitational waves are the same as in GR and vector perturbations are stable.

In spin--0 sector in the gauge $ E = 0 $ EqM read 
\begin{subequations} 
\begin{align}
(00)&: ~~~ 3H(2H+3v) \Phi + 27Hv\Psi + 3(u' + 3Hu + a^{-2} \partial_i u_i) - 6H\Psi' + 2a^{-2}\triangle \Psi = 0 \,, \\
(0i)&: ~~~~~~~~~~~~~~~~~~~~~~~~~~~~~~~~~~~  \partial_i ( H\Phi - \Psi' ) = 0 \,, \\
(ij)&: ~~~~~~~~ \partial_i \partial_j ( \Phi + \Psi ) - \delta_{i j} \Big( \triangle( \Phi + \Psi ) + a^2( 3H(2H+3v)\Phi + \\ \nonumber
& ~~~~~~~~~~~~~~~~~~~~~~~~~~~~~~~~~~~~~ + 27Hv\Psi + 2H\Phi' - 6H\Psi' - 2\Psi'' ) \Big) = 0 \;,
\end{align}
\end{subequations}
where $ \triangle = \partial_i \partial_i $. After taking into account eq. (\ref{eq_scal_fluct}) these equations become the same as in GR. Thus scalar--affine gravity is fully equivalent to GR in the linearized limit. In particular, from $ (ij) $ equations it follows that
\begin{equation}
\Psi + \Phi = 0 \;.
\end{equation}
Hence our model passes experimental constraints on the difference of these potentials \cite{Will:2014kxa}. In particular, by introducing a point--like particle of mass $ M $ at the origin of the coordinates one recovers Newton's law. Requirement (\ref{cond_for_expand}) is then fulfilled provided that
\begin{equation}
H \ll \frac{M_{pl}^2}{M} \;,
\end{equation}
which holds in all cases of physical interest. 

\bibliography{aff_ref}

\providecommand{\href}[2]{#2}\begingroup\raggedright\begin{thebibliography}{10}

\bibitem{einstein1923theory}
A.~Einstein, \emph{The theory of the affine field},  1923.

\bibitem{carmichael1925eddington}
R.~Carmichael et~al., \emph{As eddington, the mathematical theory of
  relativity}, {\emph{Bulletin of the American Mathematical Society} {\bfseries
  31} (1925) 563--563}.

\bibitem{schrodinger1943general}
E.~Schr{\"o}dinger, \emph{The general unitary theory of the physical fields},
  in \emph{Proceedings of the Royal Irish Academy. Section A: Mathematical and
  Physical Sciences}, pp.~43--58, JSTOR, 1943.

\bibitem{schrodinger1943union}
E.~Schr{\"o}dinger, \emph{The union of the three fundamental fields
  (gravitation, meson, electromagnetism)},  in \emph{Proceedings of the Royal
  Irish Academy. Section A: Mathematical and Physical Sciences}, pp.~275--287,
  JSTOR, 1943.

\bibitem{schrodinger1956space}
E.~Schr{\"o}dinger, \emph{Space-time structure}.
\newblock Cambridge University Press, Cambridge (1950-1985), 1956.

\bibitem{schrodinger1944affine}
E.~Schr{\"o}dinger, \emph{The affine connexion in physical field theories},
  {\emph{Nature} {\bfseries 153} (1944) 572}.

\bibitem{kijowski1978new}
J.~Kijowski, \emph{On a new variational principle in general relativity and the
  energy of the gravitational field}, {\emph{General Relativity and
  Gravitation} {\bfseries 9} (1978) 857--877}.

\bibitem{ferraris1981general}
M.~Ferraris and J.~Kijowski, \emph{General relativity is a gauge type theory},
  {\emph{Letters in Mathematical Physics} {\bfseries 5} (1981) 127--135}.

\bibitem{liebscher1988purely}
D.-E. Liebscher, \emph{Purely affine theories}, {\emph{Annalen der Physik}
  {\bfseries 500} (1988) 200--204}.

\bibitem{fradkin1985quantum}
E.~Fradkin and A.~A. Tseytlin, \emph{Quantum equivalence of dual field
  theories}, {\emph{Annals of Physics} {\bfseries 162} (1985) 31--48}.

\bibitem{ortin2004gravity}
T.~Ort{\'\i}n, \emph{Gravity and strings}.
\newblock Cambridge University Press, 2004.

\bibitem{Magnano:1995pv}
G.~Magnano, \emph{{Are there metric theories of gravity other than general
  relativity?}},  in \emph{{General relativity and gravitational physics.
  Proceedings, 11th Italian Conference, Trieste, Italy, September 26-30,
  1994}}, pp.~213--234, 1995,
  \href{https://arxiv.org/abs/gr-qc/9511027}{{\ttfamily gr-qc/9511027}}.

\bibitem{minkowski1977spontaneous}
P.~Minkowski, \emph{On the spontaneous origin of newtons constant},
  {\emph{Physics Letters B} {\bfseries 71} (1977) 419--421}.

\bibitem{zee1979broken}
A.~Zee, \emph{Broken-symmetric theory of gravity}, {\emph{Physical Review
  Letters} {\bfseries 42} (1979) 417}.

\bibitem{fujii1982origin}
Y.~Fujii, \emph{Origin of the gravitational constant and particle masses in a
  scale-invariant scalar-tensor theory}, {\emph{Physical Review D} {\bfseries
  26} (1982) 2580}.

\bibitem{shaposhnikov2009scale}
M.~Shaposhnikov and D.~Zenh{\"a}usern, \emph{Scale invariance, unimodular
  gravity and dark energy}, {\emph{Physics Letters B} {\bfseries 671} (2009)
  187--192}.

\bibitem{Blas:2011ac}
D.~Blas, M.~Shaposhnikov and D.~Zenhausern, \emph{{Scale-invariant alternatives
  to general relativity}},
  \href{https://doi.org/10.1103/PhysRevD.84.044001}{\emph{Phys. Rev.}
  {\bfseries D84} (2011) 044001},
  [\href{https://arxiv.org/abs/1104.1392}{{\ttfamily 1104.1392}}].

\bibitem{kubo2019planck}
J.~Kubo, M.~Lindner, K.~Schmitz and M.~Yamada, \emph{Planck mass and inflation
  as consequences of dynamically broken scale invariance}, {\emph{Physical
  Review D} {\bfseries 100} (2019) 015037}.

\bibitem{Borisov:1974bn}
A.~B. Borisov and V.~I. Ogievetsky, \emph{{Theory of Dynamical Affine and
  Conformal Symmetries as Gravity Theory}},
  \href{https://doi.org/10.1007/BF01038096}{\emph{Theor. Math. Phys.}
  {\bfseries 21} (1975) 1179}.

\bibitem{cutler1987new}
C.~Cutler and R.~M. Wald, \emph{A new type of gauge invariance for a collection
  of massless spin-2 fields. i. existence and uniqueness}, {\emph{Classical and
  Quantum Gravity} {\bfseries 4} (1987) 1267}.

\bibitem{Sotiriou:2008rp}
T.~P. Sotiriou and V.~Faraoni, \emph{{f(R) Theories Of Gravity}},
  \href{https://doi.org/10.1103/RevModPhys.82.451}{\emph{Rev. Mod. Phys.}
  {\bfseries 82} (2010) 451--497},
  [\href{https://arxiv.org/abs/0805.1726}{{\ttfamily 0805.1726}}].

\bibitem{jimenez2018born}
J.~B. Jim{\'e}nez, L.~Heisenberg, G.~J. Olmo and D.~Rubiera-Garcia,
  \emph{Born--infeld inspired modifications of gravity}, {\emph{Physics
  Reports} {\bfseries 727} (2018) 1--129}.

\bibitem{ferraris1982unified}
M.~Ferraris and J.~Kijowski, \emph{Unified geometric theory of electromagnetic
  and gravitational interactions}, {\emph{General Relativity and Gravitation}
  {\bfseries 14} (1982) 37--47}.

\bibitem{Krasnov:2007ei}
K.~Krasnov, \emph{{Non-metric gravity: A Status report}},
  \href{https://doi.org/10.1142/S021773230702590X}{\emph{Mod. Phys. Lett.}
  {\bfseries A22} (2007) 3013--3026},
  [\href{https://arxiv.org/abs/0711.0697}{{\ttfamily 0711.0697}}].

\bibitem{Krasnov:2011pp}
K.~Krasnov, \emph{{Pure Connection Action Principle for General Relativity}},
  \href{https://doi.org/10.1103/PhysRevLett.106.251103}{\emph{Phys. Rev. Lett.}
  {\bfseries 106} (2011) 251103},
  [\href{https://arxiv.org/abs/1103.4498}{{\ttfamily 1103.4498}}].

\bibitem{Poplawski:2007rx}
N.~J. Poplawski, \emph{{Gravitation, electromagnetism and cosmological constant
  in purely affine gravity}},
  \href{https://doi.org/10.1007/s10701-009-9284-y}{\emph{Found. Phys.}
  {\bfseries 39} (2009) 307--330},
  [\href{https://arxiv.org/abs/gr-qc/0701176}{{\ttfamily gr-qc/0701176}}].

\bibitem{Poplawski:2012bw}
N.~Poplawski, \emph{{Affine theory of gravitation}},
  \href{https://doi.org/10.1007/s10714-013-1625-7}{\emph{Gen. Rel. Grav.}
  {\bfseries 46} (2014) 1625},
  [\href{https://arxiv.org/abs/1203.0294}{{\ttfamily 1203.0294}}].

\bibitem{Castillo-Felisola:2015cqa}
O.~Castillo-Felisola and A.~Skirzewski, \emph{{Einstein’s gravity from a
  polynomial affine model}},
  \href{https://doi.org/10.1088/1361-6382/aaa699}{\emph{Class. Quant. Grav.}
  {\bfseries 35} (2018) 055012},
  [\href{https://arxiv.org/abs/1505.04634}{{\ttfamily 1505.04634}}].

\bibitem{Castillo-Felisola:2019wcs}
O.~Castillo-Felisola, J.~Perdiguero, O.~Orellana and A.~R. Zerwekh,
  \emph{{Emergent metric and geodesic analysis in cosmological solutions of
  (torsion-free) Polynomial Affine Gravity}},
  \href{https://arxiv.org/abs/1908.06654}{{\ttfamily 1908.06654}}.

\bibitem{azri2018induced}
H.~Azri and D.~Demir, \emph{Induced affine inflation}, {\emph{Physical Review
  D} {\bfseries 97} (2018) 044025}.

\bibitem{azri2018cosmological}
H.~Azri, \emph{Cosmological implications of affine gravity}, {\emph{arXiv
  preprint arXiv:1805.03936} (2018) }.

\bibitem{weinberg1964photons}
S.~Weinberg, \emph{Photons and gravitons in s-matrix theory: derivation of
  charge conservation and equality of gravitational and inertial mass},
  {\emph{Physical Review} {\bfseries 135} (1964) B1049}.

\bibitem{deser1970self}
S.~Deser, \emph{Self-interaction and gauge invariance}, {\emph{General
  Relativity and gravitation} {\bfseries 1} (1970) 9--18}.

\bibitem{boulware1975classical}
D.~G. Boulware and S.~Deser, \emph{Classical general relativity derived from
  quantum gravity}, {\emph{Annals of Physics} {\bfseries 89} (1975) 193--240}.

\bibitem{wald1986spin}
R.~M. Wald, \emph{Spin-two fields and general covariance}, {\emph{Physical
  Review D} {\bfseries 33} (1986) 3613}.

\bibitem{vollick2005born}
D.~N. Vollick, \emph{Born-infeld-einstein theory with matter}, {\emph{Physical
  Review D} {\bfseries 72} (2005) 084026}.

\bibitem{nielsen1978beta}
H.~B. Nielsen and M.~Ninomiya, \emph{$\beta$-function in a non-covariant
  yang-mills theory}, {\emph{Nuclear Physics B} {\bfseries 141} (1978)
  153--177}.

\bibitem{chadha1983lorentz}
S.~Chadha and H.~B. Nielsen, \emph{Lorentz invariance as a low energy
  phenomenon}, {\emph{Nuclear Physics B} {\bfseries 217} (1983) 125--144}.

\bibitem{bednik2013emergent}
G.~Bednik, O.~Pujol{\'a}s and S.~Sibiryakov, \emph{Emergent lorentz invariance
  from strong dynamics: holographic examples}, {\emph{Journal of High Energy
  Physics} {\bfseries 2013} (2013) 64}.

\bibitem{rubakov2004lorentz}
V.~Rubakov, \emph{Lorentz-violating graviton masses: Getting around ghosts, low
  strong coupling scale and vdvz discontinuity}, {\emph{arXiv preprint
  hep-th/0407104} (2004) }.

\bibitem{dubovsky2004phases}
S.~L. Dubovsky, \emph{Phases of massive gravity}, {\emph{Journal of High Energy
  Physics} {\bfseries 2004} (2004) 076}.

\bibitem{rubakov2008infrared}
V.~A. Rubakov and P.~G. Tinyakov, \emph{Infrared-modified gravities and massive
  gravitons}, {\emph{Physics-Uspekhi} {\bfseries 51} (2008) 759}.

\bibitem{Blas:2014ira}
D.~Blas and S.~Sibiryakov, \emph{{Completing Lorentz violating massive gravity
  at high energies}}, \href{https://doi.org/10.7868/S0044451015030180,
  10.1134/S1063776115030164}{\emph{Zh. Eksp. Teor. Fiz.} {\bfseries 147} (2015)
  578--594}, [\href{https://arxiv.org/abs/1410.2408}{{\ttfamily 1410.2408}}].

\bibitem{hovrava2009quantum}
P.~Ho{\v{r}}ava, \emph{Quantum gravity at a lifshitz point}, {\emph{Physical
  Review D} {\bfseries 79} (2009) 084008}.

\bibitem{hovrava2009membranes}
P.~Ho{\v{r}}ava, \emph{Membranes at quantum criticality}, {\emph{Journal of
  High Energy Physics} {\bfseries 2009} (2009) 020}.

\bibitem{asaka2005numsm}
T.~Asaka, S.~Blanchet and M.~Shaposhnikov, \emph{The $\nu$msm, dark matter and
  neutrino masses}, {\emph{Physics Letters B} {\bfseries 631} (2005) 151--156}.

\bibitem{Greensite:1992np}
J.~Greensite, \emph{{Dynamical origin of the Lorentzian signature of
  space-time}}, \href{https://doi.org/10.1016/0370-2693(93)90744-3}{\emph{Phys.
  Lett.} {\bfseries B300} (1993) 34--37},
  [\href{https://arxiv.org/abs/gr-qc/9210008}{{\ttfamily gr-qc/9210008}}].

\bibitem{Elizalde:1992ug}
E.~Elizalde, S.~D. Odintsov and A.~Romeo, \emph{{Dynamical determination of the
  metric signature in space-time of nontrivial topology}},
  \href{https://doi.org/10.1088/0264-9381/11/4/002}{\emph{Class. Quant. Grav.}
  {\bfseries 11} (1994) L61--L68},
  [\href{https://arxiv.org/abs/hep-th/9312132}{{\ttfamily hep-th/9312132}}].

\bibitem{Carlini:1993up}
A.~Carlini and J.~Greensite, \emph{{Why is space-time Lorentzian?}},
  \href{https://doi.org/10.1103/PhysRevD.49.866}{\emph{Phys. Rev.} {\bfseries
  D49} (1994) 866--878}, [\href{https://arxiv.org/abs/gr-qc/9308012}{{\ttfamily
  gr-qc/9308012}}].

\bibitem{macdowell1977unified}
S.~W. MacDowell and F.~Mansouri, \emph{Unified geometric theory of gravity and
  supergravity}, {\emph{Physical Review Letters} {\bfseries 38} (1977) 739}.

\bibitem{chamseddine1978massive}
A.~H. Chamseddine, \emph{Massive supergravity from spontaneously breaking
  orthosymplectic gauge symmetry}, {\emph{Annals of Physics} {\bfseries 113}
  (1978) 219--234}.

\bibitem{chamseddine1977supergravity}
A.~H. Chamseddine and P.~C. West, \emph{Supergravity as a gauge theory of
  supersymmetry}, {\emph{Nuclear Physics B} {\bfseries 129} (1977) 39--44}.

\bibitem{stelle1980spontaneously}
K.~Stelle and P.~C. West, \emph{Spontaneously broken de sitter symmetry and the
  gravitational holonomy group}, {\emph{Physical Review D} {\bfseries 21}
  (1980) 1466}.

\bibitem{hehl1995metric}
F.~W. Hehl, J.~D. McCrea, E.~W. Mielke and Y.~Ne'eman, \emph{Metric-affine
  gauge theory of gravity: field equations, noether identities, world spinors,
  and breaking of dilation invariance}, {\emph{Physics Reports} {\bfseries 258}
  (1995) 1--171}.

\bibitem{einstein2003meaning}
A.~Einstein, \emph{The meaning of relativity}.
\newblock Routledge, 2003.

\bibitem{Julia:1998ys}
B.~Julia and S.~Silva, \emph{{Currents and superpotentials in classical gauge
  invariant theories. 1. Local results with applications to perfect fluids and
  general relativity}},
  \href{https://doi.org/10.1088/0264-9381/15/8/006}{\emph{Class. Quant. Grav.}
  {\bfseries 15} (1998) 2173--2215},
  [\href{https://arxiv.org/abs/gr-qc/9804029}{{\ttfamily gr-qc/9804029}}].

\bibitem{Dadhich:2010xa}
N.~Dadhich and J.~M. Pons, \emph{{On the equivalence of the Einstein-Hilbert
  and the Einstein-Palatini formulations of general relativity for an arbitrary
  connection}}, \href{https://doi.org/10.1007/s10714-012-1393-9}{\emph{Gen.
  Rel. Grav.} {\bfseries 44} (2012) 2337--2352},
  [\href{https://arxiv.org/abs/1010.0869}{{\ttfamily 1010.0869}}].

\bibitem{eisenhart2012non}
L.~P. Eisenhart, \emph{Non-Riemannian geometry}.
\newblock Courier Corporation, 2012.

\bibitem{borchsenius1976extension}
K.~Borchsenius, \emph{An extension of the nonsymmetric unified field theory},
  {\emph{General Relativity and Gravitation} {\bfseries 7} (1976) 527--534}.

\bibitem{Poplawski:2007ik}
N.~J. Poplawski, \emph{{A Unified, purely affine theory of gravitation and
  electromagnetism}},  \href{https://arxiv.org/abs/0705.0351}{{\ttfamily
  0705.0351}}.

\bibitem{Krasnov:2017epi}
K.~Krasnov and R.~Percacci, \emph{{Gravity and Unification: A review}},
  \href{https://doi.org/10.1088/1361-6382/aac58d}{\emph{Class. Quant. Grav.}
  {\bfseries 35} (2018) 143001},
  [\href{https://arxiv.org/abs/1712.03061}{{\ttfamily 1712.03061}}].

\bibitem{Kharuk:2018ums}
N.~V. Kharuk, S.~N. Manida, S.~A. Paston and A.~A. Sheykin, \emph{{Modifying
  the theory of gravity by changing independent variables}},
  \href{https://doi.org/10.1051/epjconf/201819107007}{\emph{EPJ Web Conf.}
  {\bfseries 191} (2018) 07007},
  [\href{https://arxiv.org/abs/1811.00831}{{\ttfamily 1811.00831}}].

\bibitem{peeters2018cadabra2}
K.~Peeters, \emph{Cadabra2: computer algebra for field theory revisited.},
  {\emph{Journal of open source software.} {\bfseries 3} (2018) 1118}.

\bibitem{peeters2007introducing}
K.~Peeters, \emph{Introducing cadabra: A symbolic computer algebra system for
  field theory problems}, {\emph{arXiv preprint hep-th/0701238} (2007) }.

\bibitem{Will:2014kxa}
C.~M. Will, \emph{{The Confrontation between General Relativity and
  Experiment}}, \href{https://doi.org/10.12942/lrr-2014-4}{\emph{Living Rev.
  Rel.} {\bfseries 17} (2014) 4},
  [\href{https://arxiv.org/abs/1403.7377}{{\ttfamily 1403.7377}}].

\end{thebibliography}\endgroup

\end{document}